# THE INCREASED NEED FOR FCC MERGER REVIEW IN A NETWORKED WORLD
by Harold Feld[*]

## INTRODUCTION

As concentration in the media has grown, so too has the debate over the role of the FCC in merger review. On the one hand, a host of critics has disparaged FCC review of mergers as redundant of the more expert antitrust agencies, costly, time consuming, a means by which the agency expands its powers beyond either statutory or prudential limits, and a general hindrance to the development of an efficient market in the "new economy" where the old regulatory paradigms do not apply. [CITES] Others have defended the FCC's merger review as an important regulatory tool and a vital protection to the "marketplace of ideas" in a world dominated by an ever-smaller number of conglomerates combining proprietary content with proprietary platforms. [CITES]

Most recently, the new Chairman of the FCC Michael Powell, joined by the two other republican Commissioners, stated his formulation of the FCC's public interest standard in merger reviews. [CITE FOX/Chris-Craft] Henceforth, the FCC will distinguish between media mergers and common carrier mergers and apply different standards to the two. The FCC will review media mergers to see whether they comply with the FCC's existing rules. If the proposed media merger violates no rule, no further inquiry is necessary. Only if the proposed media merger violates an FCC rule will the Commission look further, to see if some waiver or condition is necessary. By contrast, in common carrier mergers, the FCC will continue to apply the more general public interest test set forth in the Commission's Order reviewing the Bell Atlantic/Nynex merger. [CITE]. Chairman Powell, who as a Commissioner frequently inveighed against imposing conditions in media mergers, [CITES], justified this dichotomy on the grounds that the Commission had extensive structural rules in media regulation, and that further review in the merger context would "eviscerate" the benefit of having such rules.[1]

---

[*]Associate Director, Media Access Project.

Chairman Powell's view of the FCC's role in merger review presents difficulties, not the least of which being that the neat division of the world into common carriers and mass media outlets no longer exists. The much discussed and anticipated convergence of technologies wedding content and conduit renders this neat division obsolete on arrival. It also fails to account for the increasingly powerful network effects and other barriers to entry created by such mergers that inhibit or prevent competition from new entrants to offset the increased concentration. It also ignores the plain reading of the communications Act, which requires a specific inquiry into each license transfer as to whether it serves the public interest. [CITE]

This most profoundly effects the diversity of voices and opinions available in the marketplace of ideas. As the dissenters in the Fox/Chris-Craft merger observed, the ability to pass information freely and to hear a diversity of views is the essential machinery of democracy. The FCC's abandonment of its responsibilities allows the few controllers of vast communications networks to act as media gate keepers, controlling what the public can see, hear, and say. Given the assault upon the FCC's structural rules by the District of Columbia Circuit, and the continued concentration of both content production and distribution channels in the hands of a few under the existing regime, the FCC's new standard of review constitutes a grave threat to the freedom of the marketplace of ideas.

I have written elsewhere in defense of the FCC's merger review authority and how it differs from that of the antitrust agencies – the Department of Justice (DOJ) and the Federal Trade

---

[1] The two democratic Commissioners, dissented from Chairman Powell's narrow vision of the public interest.

Commission (FTC) – charged with reviewing mergers generally under the Hart-Scott-Rodino Act (HSRA).[2] This paper focuses on the flaws in the Chairman's new formulation and the need for the FCC to perform a more thorough merger review taking into account the reality of the new, networked world. Contrary to myth, absent federal regulation, the networked world is marked by high barriers to entry and little opportunity for competition to develop. In particular, the marketplace of ideas remains highly vulnerable to mergers that do not pose significant competitive risk under traditional antitrust analysis.

Looking briefly to other areas where Congress has imposed an additional layer of merger review by an expert agency, notably electric power & natural gas, banking, and transportation, a pattern emerges. These areas are generally marked by centrality to the economy, high barriers to entry, strong network effects, and the presence of local markets that would go unserved or underserved without federal intervention. Because of these factors, Congress has found these areas affected with the public interest. Because of the dangers presented by concentration in these industries, Congress has proven far more "risk averse" in allowing consolidation. Congress therefore created a requirement that parties wishing to merge demonstrate the merger serves the public interest, a more stringent requirement than that generally applicable under Section 7 of the Clayton Act.[3]

It is instructive to observe that in areas partially or entirely deregulated by Congress, transportation and banking, barriers to entry had generally declined prior to deregulation and Congress left intact laws requiring nondiscriminatory service or other rules designed to address the needs of potentially underserved communities.

---

[2] Pub. L. 94-435, 90 Stat. 1394 (1976) (codified at 15 U.S.C. §18a).

[3] Section 7 of the Clayton Act (codified at 15 U.S.C. §18) prohibits mergers or acquisitions "the effect of such acquisition may be substantially to lessen competition, or to tend to create a monopoly."

Turning to communications, the factors arguing for continued review remain. Telecommunications remains not merely vital to the economy, but to the proper functioning of democracy. In the absence of federal rules mandating non-discrimination network effects and high costs operate to exclude new entrants. For example, before the Cable Act of 1992 and the Satellite Home Improvement Act of 1999 provided DBS broadcasters with access to national and local programming, DBS could not compete with cable in the program distribution market. In the absence of an open access requirement for cable, cable has remained impenetrable to local (and most national) ISPs.

Contrary to the Chairman's formulation, however, the potentials for abuse are not wholly addressed by either the existing rules or by the rulemaking process. Mergers permit new configurations of content and conduit not previous subject to rulemaking consideration. The merger process is designed to give these new configurations, which effect not merely the participants but the nation as a whole, the regulatory scrutiny they require. In addition, the flexibility of merger review permits the FCC to narrowly tailor any needed regulation of the needs of the marketplace. Nor does it "eviscerate" the value of the FCC's existing rules. Finally, in light of the recent hostility of the Federal Court of Appeals of the District of Columbia Circuit to the FCC's structural rules, intense scrutiny of mergers may become the FCC's only means of preserving what diversity remains in the marketplace of ideas.

## PART I – OTHER AREAS SUBJECT TO ADDITIONAL REVIEW

In only a few sectors of the economy has Congress chosen to impose an additional layer of review of mergers beyond that mandated by the HSRA. Under the Bank Merger Act, mergers of banks and other savings institutions are reviewed by the agency charged with regulating the savings institution.[4] Mergers of electric utilities and natural gas providers are reviewed by the Federal Energy Regulatory Commission (FERC).[5] Transportation mergers, first regulated by the Interstate Commerce Commission in 1887 and focusing at that time on regulation of the railroads,[6] were deregulated by Congress over time, culminating in the abolishment of the ICC in 1995.[7]

These areas have several elements in common. First, they are traditionally marked by large networks with significant build out costs, where the combination of network effects and other barriers to entry create the potential for monopoly and anticompetitive practices. In the case of electricity and natural gas, this arises from the classic conditions that used to define a "natural monopoly." The movement of natural gas or electricity is relatively cheap once the network is in place, but build out costs are high. Larger networks allow much greater efficiencies with little marginal costs for each new customer, so that larger networks can simply underprice new entrants until they fold. When the ICC was first established, railroads operated under similar conditions, as did the airline industry in its infancy when start up costs were enormous.[8] Banks, while not have the same physical access problem, require enormous start up capital and new entrants suffer from a lengthy period when they must establish their trustworthiness to the market.

---

[4] *See generally* PHILIP E. AREEDA & HERBERT HOVENKAMP, ANTITRUST LAW: AN ANALYSIS OF ANTITRUST PRINCIPLES AND THEIR APPLICATION (2000) §2C-2 ¶251f ("Areeda"). As Areeda explains, different savings institutions are regulated by different federal agencies.

[5] Areeda ¶251g.

[6] 24 Stat 379 (1887). In many ways, the ICC served as a model for the FCC, including its use of the public interest standard. [CITE]

[7] ICC Termination Act of 1995, Pub. L. No. 104-88, *see also* 49 USC §§701 *et seq.* (transferring remaining powers to newly created Surface Transportation Board).

[8] The same case cannot be said for motor transport, where few barriers to entry exist. It seems likely that regulation of motor transport is something of an aberration, and as Areeda suggests, probably should not have been subject to merger regulation at all. Areeda ¶251i.

Because circumstances make it difficult for new entrants, the market tends toward concentration with the inevitable problems of monopoly control. This problem is further complicated by recognition of these services as critical to the average citizen to participate in society. Yet without federal intervention, the poorest members of society and minorities traditionally subject to discrimination may not receive service, or may receive only inferior service at high prices.

As a result, Congress has found that these industries are affected with the public interest.[9] One aspect of regulation in the public interest is that mergers must therefore serve the public interest.[10] Contrary to critics, then, the treatment of the merger process as a quasi-regulatory adjudicative proceeding does not represent some sort of agency bias toward expanding its own power or toward regulation generally. [CITES] Rather, it represents a reflection of Congress' affirmative charge to the agency.

It is instructive to examine the areas Congress has deregulated. Although Congress initially regulated transportation tightly, it has deregulated it almost completely. Looking to the reality of transportation, the reasons conform to the general model set forth above. Rail, the mode of transportation requiring the greatest build out, has ceased to be a dominant form of transportation. Travel by car and transportation by truck is not marked by the same barriers to entry as rail travel.

---

[9]*See, e.g.*, 12 U.S.C. §2901 (banks managed "for the needs of the community"); 15 U.S.C. 717(a) (distribution of natural gas is "affected with the public interest").

[10]*See, e.g.*, 16 U.S.C. 824b (FERC shall not permit the transfer of assets unless it finds the transfer "consistent with the public interest."

Even barriers to entry in air travel and freight have greatly diminished.[11] As a result, the need for government management of the industry has disappeared and review of mergers under the Clayton Act standard suffices to preserve competition.

## PART II -- THE COMMUNICATIONS MARKET

Looking to the communications market, the factors that mandate regulation – and therefore regulation of mergers – in the public interest remain. Despite recent set backs to the technology sector, telecommunications and mass media remain critical components of participation in modern society. Most significantly, the mass media remain the primary means by which Americans educate themselves on news and access entertainment. Even if one includes the Internet as an alternative to mass media such as television, radio and cable, that merely pushes the problem back to the telecommunications sector as a whole.

Indeed, with the continued convergence of all telecommunications sectors, the division of the world into convenient categories such as "mass media" and "common carrier" become increasingly artificial. As cable operators increasingly offer the next generation of interactive services such as broadband access and interactive television and television broadcasters use their spectrum to offer telecommunication services such as datacasting, the more a merger analysis based solely on the historic function of the licensees misses the full story.[12] The Commission is charged with monitoring

---

[11]In no small part because the necessary infrastructure, such as roads, air ports, and traffic management, are non-proprietary.

[12]As the Fox/Chris-Craft dissenters pointed out, the agency itself has previously recognized the need for broader application of the public interest test in mass media mergers. Indeed, in the AOL-TW Merger, the Commission explicitly rejected the attempt by the applicants to portray the merger as a simple cross-check against the Commission's rules. [CITE Wright Letter]

and encouraging competition and diversity in the Communications sector as a whole, it cannot reasonably limit its analysis to one historic activity of the licensee and consider its statutory obligation filled.

Critics frequently charge that the drafters of the Telecommunications Act of 1996 demonstrated a preference for deregulation, including a preference for more relaxed merger scrutiny. This confuses a preference for *competition* with a preference for *deregulation*. The drafters of the 1996 Act expected a competitive environment to emerge which would render structural regulation unnecessary. But -- in an apparent recognition that even Congress is not omniscient -- the drafters did not curtail the FCC's function as the monitor of the telecommunications sector or restrain its merger review authority. Indeed, the alteration Congress did make to the FCC's merger authority, removing the ability of the FCC to immunize certain mergers from antitrust review, suggests the opposite conclusion: Congress knew that its predictive judgment might prove wrong, and did not wish to remove the one remaining safety valve to industry concentration.

Events have born out the wisdom of this prudence. The changes created by the 1996 Act have not created the anticipated competition. The seven regional Bell operating companies and GTE have merged down to four expanded local monopolies with no interest in competing in each other's territories. Cable operators have likewise consolidated, producing neither overbuilders nor a second telephone wire into the home. Radio and television broadcasters have also consolidated, with two television networks reaching more than 40% of the total audience of the United States in violation of the FCC's existing rules and the remaining networks at the limit.

The difficulties of would-be entrants demonstrate that the physical and economic barriers to entry remain in the telecommunications sector. In the provision of telecommunications and Internet services, an initially vigorous attempt at overbuilding has collapsed, leaving the local monopolies to dominate the market. In addition to the expense and difficulty in wooing customers from established networks, competitors have cited the control of the local network by a competitor and the failure of the FCC to enforce compliance with the regulations mandating access as a primary reason for the failure of competition to emerge. In the cable industry, overbuilders and alternate multichannel video programming distributors (MVPDs) have run into similar problems, facing resistance from entrenched cable operators. Indeed, only one competitor to cable has emerged in the MVPD market, Direct Broadcast Satellite (DBS). This success was not achieved by competition in the free market, but by massive intervention on behalf of DBS on the part of Congress to provide access to programming and require non-discrimination by vertically integrated cable operator.

Finally, in the broadcast medium, the continued scarcity of licenses -- re-enforced by the ability of the incumbents to lobby effectively against the FCC's attempts to open the spectrum to new entrants -- and the relaxation of the FCC's structural rules has lead to consolidation on an unprecedented level. The relaxation of the "dual network rule" allows the larger established networks and their emerging competitors to exist under joint ownership, while the relaxation of the national ownership limits and the one-to-a-market rule have allowed networks and large ownership groups to acquire the majority of independent broadcasters or smaller groups which might provide local diversity and competition for advertising revenue.

Nor has competition emerged from the Internet. Mistaking the FCC's mandate of non-discrimination for "enhanced services" for lack of regulation, Congress proudly proclaimed in the 1996 Act that "\_\_\_\_\_." Those opposed to regulation frequently cite the Internet as the paradigm for "unregulation" and competition.

This view, of course, overlooks or ignores the lengthy history of government-mandated open access in the pre-1996 Act world. As others have observed, the competitive and open Internet that existed at the time of the 1996 Act did not happen by accident or by a miracle of the invisible hand of the market. It was the direct result of a *regulatory* decision to promote competition via regulation by mandating open access to the communications network.

This revisionism has consequences. The FCC has refused to mandate the same non-discrimination requirements on the emerging high-speed networks, with the exception of the high-speed telephone networks, where non-discriminatory access to the physical network is required by law.

As a result of this failure to regulate, competition to existing MVPDs or even basic telecommunications services has failed to emerge from the Internet. Congress' failure to provide a mechanical license for Internet broadcasting, as it did with cable an DBS, has effectively prevented the Internet from emerging as a competitor to established radio and television broadcasters or cable. Those anticipating that the Internet would provide a competitive market in telephony services overlooked both the technical challenges and the resistance of the local monopolies. Furthermore, the failure to impose open access on cable, the emerging leader in residential broadband access, has lead to a steady decline in competition in the provision of even basic Internet services. Significantly, the one area that has hitherto been marked by the greatest competition, commercial Internet access, exists on the most regulated and mandatorily opened network.

To summarize, the factors that mandate for a public interest merger review in the telecommunications sector remain. In the absence of government action, competition has not and will not emerge. As in other sectors identified in Part I, the result of concentration in the absence of government regulation would be to leave segments of society vulnerable to total lack of service or significant underservice, and to subject society as a whole to the ills of monopoly control over a vital sector of the economy.

Added to this factor lies the profound effect of such concentration on democracy. Since the first days of radio, Congress, the FCC and the courts have recognized that the operation of the mass media have a profound effect on the free flow of ideas and information. Where one entity or oligopoly control that flow, it creates the risk that the "media gatekeeper" will control what the public see, hear and say. The dangers of such a system of centralized control of information are no less troublesome when a private corporation exercises that power as when a government censor does.[13]

## PART III -- MERGER REVIEW RATHER THAN STRUCTURAL REGULATION

While the arguments above may make the case for regulation generally, they do not necessarily make the case for merger review. It would seem, at first glance, that to the extent a need for regulation remains, Chairman Powell's formulation in *Fox/Chris-Craft* has merit.

A closer examination, however, reveals the flaws in the Chairman's formulation. Briefly, the rules establish guidelines, but these guidelines cannot take the place of a fact-based determination in complex mergers -- particularly in the emerging networked world.

---

[13]Even if one accepts the argument that a monopolist has no incentive to discriminate and every incentive to maximize programming diversity, it makes monopoly control no more palatable. Leaving aside the numerous real-world examples in which monopolists have chosen to forgo short-term profits from other sources in favor of maintaining their core monopoly, the *ability* of a single entity to small group to act as a censor to the public creates problems with the free flow of information and ideas. As an initial matter, society should not have to tolerate in a private entity what it would find intolerable on government. Furthermore, the willingness of information providers to self-censor and tailor offerings to the perceived preferences of the media gatekeeper creates as much of a bottleneck to information as active discrimination. Finally, the efficiencies achieved by monopoly may act against the public interest. It is efficient to scale news resources so that a single newsroom or news network provides news, while the remaining channel on the network provide entertainment. But the public suffers from the failure of competition in the news sector. The number of editorial voices interpreting the news is diminished and fewer items will be covered. As a result, access to information, a critical component of a democratic society, is curtailed.

Indeed, the dynamic nature of the networked world extolled by the critics of merger review is precisely what makes merger review superior to rulemaking. A rulemaking proceeding cannot foresee all possibilities. When a new situation emerges, the Communications Act and prudence both demand an examination to determine whether the transfer of an FCC license serves "the public interest, convenience and necessity." The failure to foresee a possible combination should not act as a free pass from public scrutiny.

Consider, for example, the merger of America Online and Time Warner. No FCC rule explicitly addressed the situation that arose when the largest Internet access provider and one of the most aggressive proponents of non-cable broadband bought a vertically integrated cable company. Under the Chairman's formulation, the FCC should simply have approved the merger without examination of the consequences. In the Powell formulation, the FCC should simply permit the merger to take place and any subsequent injury to competition or the marketplace of die as should be addressed via the rulemaking process.